\documentclass[aps,amsmath,twocolumn]{revtex4}

\usepackage{amsmath}
\usepackage{graphicx}

\begin{document}

\title{Slow and fast light dynamics in a chiral cold and hot atomic
medium }
\author{Bakht A Bacha}
\affiliation{Department of Physics, Hazara University, Pakistan}
\author{Fazal Ghafoor}
\affiliation{Department of Physics, COMSATS Institute of
Information Technology, Islamabad, Pakistan}
\author{Rashid G Nazmidinov}
\affiliation{Laboratory of Theoretical Physics JINR, Moscow
Region, Dubna, Russia}

\begin{abstract}
We study Chiral Based Electromagnetically Induced Transparency
(CBEIT) of a light pulse and its associated subluminal and
superluminal behavior through a cold and a hot medium of 4-level
\textit{double-Lambda type} atomic system. The dynamical behavior
of this chiral based system is temperature dependent. The magnetic
field based chirality and dispersion is always opposite as
compared with the electric field ones. Contrastingly, the response
of the chiral effect along with the incoherence Doppler broadening
mechanism enhances the superluminal behavior as compared with its
traditional degrading effect. Nevertheless, the intensity of a
coupled microwave field destroys the coherence of the medium and
degrade superluminality and subluminality of the sysmtem. The
undistorted retrieved pulse from a hot chiral medium delays by
$896 ns$ than from a cold chiral medium under same set of
parameters. Nevertheless, it advances by $-31n s$ in the cold
chiral medium when a suitably different spectroscopic parameters
are selected. The corresponding group index of the medium and the
time delay/advance, are studied and analyzed explicitly [Note: A
revise version is under preparation]
\end{abstract}
\maketitle

Quantum coherence is widely initiated through strong laser fields
in multi-level atomic or molecular system interacting with
electromagnetic fields. The coherent atomic or molecular states,
which are prepared by the laser field, may then generates
phenomenon of quantum interference among various transition
amplitudes. The control over the response function of the media is
then the result of the control of quantum coherence and
interference effects. In the recent two decades,
Electromagnetically induced transparency EIT and related quantum
interference effect in various atomic schemes has been studiedg
theoretically as well as experimentally \cite{fim2005}. The
modification of optical properties of quantum material media
remained a hot area due its expected large number of useful
applications \cite{MD2001,rw2009}.
\begin{figure}[t]
\centering
\includegraphics[width=3.5in]{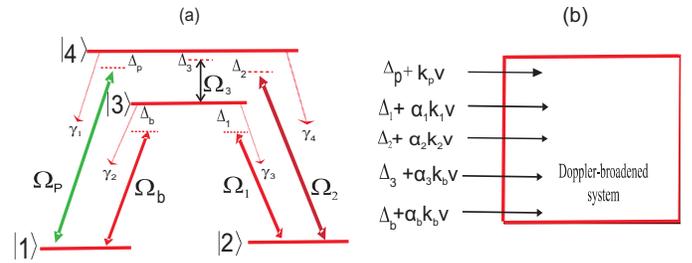}
\caption{(a) Schematics of  the Double Lambda Atomic system. (b)
Doppler-broadened system} \label{figure1}
\end{figure}

One of the modified media is the Left-handed medium which posses
simultaneously negative electric permittivity and magnetic
permeability \cite{pendry1996,pendry1998,pendry1999}. Earlier
explored left-handed media were anisotropic in nature.
Nevertheless, later with the use of new techniques isotropic left
handed media were also realized \cite{shelby2001, shelbyrd2001}.
Both the permittivity and permeability were required negative
($\epsilon,\mu <0$) for a negative refractive index with the
magmatic dipole moments very smaller than the electric one.
Nevertheless, due to the above rigorous condition, media with
negative permeability in the optical frequencies domain met very
rarely. Even with these complications, material media of negative
refractive index based on EIT and photonic resonance were also
studied.

Obviously, the quantum coherence and interference effect has
remarkable impact on the speed of a probe pulse. Consequently,
high degree control over the group velocity has been made possible
both theocratically and experimentally. The detail study in this
direction is very large. Nevertheless, some related works can be
found in
\cite{he1991,AK95,LV99,DD99,OS96,cg1970,scw1982,bs1985,Bs2005,yrc2002,bab2013,MM99,AK95,gs03,shang08}
and in the references therein. Further, Moti and his co-worker
have been presented an experimental demonstration for temporal
cloaking while using the concepts of time-space duality between
diffraction and dispersive broadening. Their experiment may be
significant effort toward the development of complete
spatio-temporal cloaks while using negative refractive indices
\cite{Moti2012}. Glasser \textbf{et al} \cite{RYANT2012} used
four-wave mixing in hot atomic vapors and experimentally measured
multi-spacial-mode images with the use of negative group velocity.
In this experiment the degree of temporal reshaping was quantified
and increased with the an increase of the pulse-time-advancement.
Furthermore, to impart the information of images in an optical
pulse propagating through a region of anomalous dispersion may
have a large number of potential applications while avoiding the
pulse reshaping. Four-waves mixing in double-Lambda scheme of the
Rubidium atom has also been shown to exhibit multi-spatial-mode
entanglement \cite{vbmp2008}. Nevertheless, a system exhibiting
less losses or gain may lead to an ideal multi-spatial-mode
entanglement. The spatial properties of the multi-spatial-mode
entanglement may also be precessed in the presence of fast light
medium if the system is constrained to less losses or gains.

In this paper we investigate the effect of a chiral cold and hot
medium of double Lambda type atoms on a propagating pulse. The
chiral medium is a medium in which the electric polarization is
coupled to the magnetic field component of the incident
electromagnetic fields in free space while the magnetization is
coupled to the incident electric field components. These
mechanisms of the atom-field interaction ensures the proposed
medium to posses the characteristics of negative refractive index
even with no need of negative permittivity and permeability
\cite{jbpendry2004}. In this connection, we select a chiral cold
and hot atomic medium and investigate the effect of chirality on
EIT and consequently on subluminal and superluminal behavior of
the propagating pulse. Unlike the traditional degrading behavior
of the Doppler broadened medium, here we explored contrastingly,
an enhancement in the superluminal behavior of the propagating
probe pulse for the hot chiral medium. Furthermore, we note that
the nature of the propagation is different for suitably different
set of parameters. Our explored undistorted retrieved pulse from a
hot chiral medium delays significantly as compared with a cold
chiral medium with a similar conditions. Nevertheless, the pulse
advances enormously with a suitably different spectroscopic
parameters. Evidently, the behavior of the propagating pulse is
modified when the medium is reverted from Doppler broadened (hot)
to the Doppler-free (cold) mode and vice versa. The group index of
the medium, time delay/advance is also studied and analyzed
explicitly.

We choose a four-level atomic system in \textit{Double-lambda}
configuration as shown in Fig. 1. The lower hyperfine ground
levels $\left\vert 1\right\rangle $ and $\left\vert 2\right\rangle
$ are coupled with the upper excited level $ \left\vert
4\right\rangle $, by a control field having Rabi frequency
$\Omega_2$, and with a probe field having Rabi frequency
$\Omega_p$, respectively. The excited hyperfine state $ \left\vert
3\right\rangle $ is coupled with the same lower ground levels by a
magnetic field with the Rabi frequency $\Omega _{b}$, and a
control field with the Rabi frequency $\Omega _{1}$, respectively.
We also coupled the two excited hyperfine levels $\left\vert
3\right\rangle$ and $\left\vert 4\right\rangle$ by a microwave
field with the Rabi frequency $\Omega_3$. Further, to keep the
nature of interaction general we consider these fields detuned
from their respective coupling energy levels and are defined as
under: $\Delta_1=\omega_{13}-\omega_1$,
$\Delta_2=\omega_{14}-\omega_2$ and
$\Delta_p=\omega_{14}-\omega_p$, $\Delta_b=\omega_{13}-\omega_b$,
$\Delta_3=\omega_{34}-\omega_3$. Next to drive the equations of
motion and analyze optical response functions for the system, we
proceed with the following interaction picture Hamiltonian in the
dipole and rotating wave approximations as:
\begin{eqnarray}
H(t)& =&-\frac{\hbar }{2}\Omega _{p}\exp[-i\Delta_pt] \left\vert
1\right\rangle \left\langle 4\right\vert-\frac{\hbar }{2}\Omega
_{b}\exp[-i\Delta_bt ]\left\vert 1\right\rangle \left\langle
3\right\vert\nonumber\\&& -\frac{\hbar }{2}\Omega _{1}[-i\Delta_1
t]\left\vert 2\right\rangle \left\langle 3\right\vert-\frac{\hbar
}{2}\Omega _{2}\exp[-i\Delta_2 t]\left\vert 2\right\rangle
\left\langle 4\right\vert\nonumber\\&& -\frac{\hbar }{2}\Omega
_{3}\exp[-i\Delta_2 t+i\varphi]\left\vert 3\right\rangle
\left\langle 4\right\vert+H.c.
\end{eqnarray}%
The general form of density matrix equation is given by the
following relation:
\begin{equation}
\frac{d\\\rho_t}{dt}=\frac{-i}{\hbar}[
\rho_t,H_t]-\frac{1}{2}\Gamma_{ij} \sum(  \sigma^\dagger  \sigma
\rho+\rho \sigma^\dagger \sigma-2\sigma \rho  \sigma^\dagger)
\end{equation}
where $\sigma^\dagger$ is the raising operator and $\sigma$ is
lowering operator for the four decays processes. Here, in the
dynamical equation we use the transformation equation of fast
varying transition matrix elements to the slowly varying
transition matrix element through the
$\rho_{ij}=\widetilde{\rho}_{ij}\exp[-i\Delta_{ij}]$, to remove
the time dependent exponential factors. Finally, we obtained the
three coupled dynamical equations for our system as:
\begin{eqnarray}
\overset{\cdot }{\overset{\sim }{\rho
}}_{14}&=&[i\Delta_p-\frac{1}{2}(\gamma_1+\gamma_2)]\widetilde{\rho}_{14}
+\frac{i}{2}\Omega_p(\widetilde{\rho}_{11}-\widetilde{\rho}_{44})\nonumber\\&&+\frac{i}{2}\Omega_3\exp[i\varphi]
\widetilde{\rho}_{13}+\frac{i}{2}\Omega_2 \widetilde{\rho}_{12}
-\frac{i}{2}\Omega_b \widetilde{\rho}_{34},
\end{eqnarray}
\begin{eqnarray}
\overset{\cdot }{\overset{\sim }{\rho
}}_{13}&=&[i\Delta_b-\frac{1}{2}(\gamma_1+\gamma_2)]\widetilde{\rho}_{13}
+\frac{i}{2}\Omega_b\widetilde{\rho}_{11}+\frac{i}{2}\Omega_1
\widetilde{\rho}_{12}\nonumber\\&&-\frac{i}{2}\Omega_b
\widetilde{\rho}_{33}-\frac{i}{2}\Omega_p
\widetilde{\rho}_{43}+\frac{i}{2}\Omega_3\exp[-i\varphi]
\widetilde{\rho}_{14},
\end{eqnarray}
\begin{eqnarray}
\overset{\cdot }{\overset{\sim }{\rho
}}_{12}&=&[i(\Delta_p-\Delta_2)-\frac{1}{2}(\gamma_1+\gamma_2+\gamma_3+\gamma_{4})]\widetilde{\rho}_{12}
-\frac{i}{2}\Omega_1\widetilde{\rho}_{13}\nonumber\\&&+\frac{i}{2}\Omega_2
\widetilde{\rho}_{14}-\frac{i}{2}\Omega_b
\widetilde{\rho}_{32}-\frac{i}{2}\Omega_p \widetilde{\rho}_{42},
\end{eqnarray}
In the derivation of the above dynamical equation we considered
$\Omega_p$, and $\Omega_b$ in the first order, while $\Omega_1$,
$\Omega_2$ and $\Omega_3$ are assumed in all order of the
perturbations. Next, we assume the atoms initially in the ground
state $\left\vert 1\right\rangle$. Therefore, the population
initially in the other states is zero i.e.,
$\widetilde{\rho}_{11}=1$,
$\widetilde{\rho}_{44}=\widetilde{\rho}_{34}=0$,
$\widetilde{\rho}_{32}=\widetilde{\rho}_{33}=0$,
$\widetilde{\rho}_{42}=\widetilde{\rho}_{43}=0$. Next, we assume
the temperature of the medium hot and assume the system
Doppler-broadened. To incorporate the broadening effect in the
system we replaced the detuning parameters by: $\Delta_1=\Delta_1+
\alpha_1 k_1 v$, $\Delta_2=\Delta_2+ \alpha_2 k_2 v$,
$\Delta_b=\Delta_b+ \alpha_3 k_b v$ $\Delta_p=\Delta_p+ k v$. In
the above replacement If we consider $\alpha_{i=1,2,3}=1$, then
the coherent fields are co-propagating with the probe field while
$\alpha_{i=1,2,3}=-1$ represent their corresponding counter
propagation. Here $k_1$, $k_2$, $k_3$ ,$k_p$, $k_b$ are the wave
vectors of coherent fields and the probes electric and magnetic
fields. Nevertheless, we assumed $k_1=k_2=k_3=k_b=k_p=k$, in our
analysis for a simplicity. To evaluate
$\widetilde{\rho}_{14}^{(1)}$ and $\widetilde{\rho}_{13}^{(1)}$ in
its steady state limit we used the following expression.
\begin{equation}
Y(t)=\int^{t}_{-\infty}e^{-M(t-t^,)}Xdt^,=-M^{-1}X,
\end{equation}
where $Y(t)$ and $X$ are column matrices while M is a 3x3 matrix.
The solutions obtained are given bellow
\begin{eqnarray}
\widetilde{\rho}_{14}^{(1)}=\Omega_p\beta_{EE}+\Omega_b\beta_{EB},
\end{eqnarray}
\begin{eqnarray}
\widetilde{\rho}_{13}^{(1)}=\Omega_p\beta_{BE}
+\Omega_b\beta_{BB},
\end{eqnarray}
where
\begin{eqnarray}
\beta_{EB}=\beta_{BE}^*=&\frac{-(i\Omega_1\Omega_2+2\Omega_3A_3\exp[i\varphi])}{2[\Omega_1\Omega_2\Omega_2\sin\varphi
-\Omega^2_1A_1+\Omega^2_2A_2+(\Omega^2_3+4A_1A_2)A_3]},
\end{eqnarray}
and
\begin{eqnarray}
\beta_{EE,(BB)}=&\frac{(-i)i(\Omega^2_1(+)-4A_{2(1)}A_3)}{2[\Omega_1\Omega_2\Omega_2\sin\varphi
-\Omega^2_1A_1+\Omega^2_2A_2+(\Omega^2_3+4A_1A_2)A_3]},
\end{eqnarray}
while
\begin{equation}
A_1=[i(\Delta_p+kv)-\frac{1}{2}(\gamma_1+\gamma_2)],
\end{equation}
\begin{equation}
A_3=[i(\Delta_b+\alpha_3kv)-\frac{1}{2}(\gamma_1+\gamma_2)],
\end{equation}
and
\begin{equation}
A_2=[i(\Delta_1+\alpha_1kv-\Delta_b-\alpha_3kv)-\frac{1}{2}(\gamma_1+\gamma_2+\gamma_3+\gamma_4)].
\end{equation}
In Eqs. (7) and (8) $\beta_{EE}$ and $\beta_{BB}$, appear for
electric and magnetic polarizabilities, while $\beta_{EB}$ and
$\beta_{BE}$, represents the chirality coefficients. The electric
polarization is defined by $P=N\sigma_{14}\rho_{14}$, and
magnetization can be measured from $M=N\mu_{13}\rho_{13}$, where
$\sigma_{14}$, is the electric and $\mu_{13}$, is the magnetic
dipole moments and $N$, represents atomic number density. The Rabi
frequencies are related to electric and magnetic fields through
the relations, $\Omega_p=\sigma_{14}.E/\hbar$ and
$\Omega_b=\mu_{13}.B/\hbar$, respectively. The electric and
magnetic polarizations collectively is given by a simplified form
as under
\begin{equation}
P,(M)=\frac{N\sigma^2_{14}\beta_{EE,(BB)}}{\hbar}E,(B)+\frac{N\sigma_{14}\mu_{13}\beta_{EB,(BE)}}{\hbar}B,(E),
\end{equation}
where $B=\mu_{0}(H+M)$. Substituting the value of $B$ in magnetic
polarization, and rearranging the equation we obtained the
magnetization as:
\begin{eqnarray}
M(kv)=\frac{N\mu_0\mu^2_{13}\beta_{BB}}{\hbar-N\mu_0\mu^2_{13}\beta_{BB}}H+\frac{N\mu_{13}
\sigma_{14}\beta_{BE}}{\hbar-N\mu_0\mu^2_{13}\beta_{BB}}E
\end{eqnarray}
Substituting the value of $M$, and $B=\mu_{0}(H+M)$, in $P$, we
obtained the expression for electric polarization as under
\begin{eqnarray}
P(kv)&=&\frac{G+N\sigma^2_{14}\beta_{EE}(\hbar-N
\mu_0\mu^2_{13}\beta_{BB})}{\hbar(\hbar-N
\mu_0\mu^2_{13}\beta_{BB})}E\nonumber\\&&+\frac{N\sigma_{14}\mu_0\mu_{13}\beta_{EB}}{(\hbar-N
\mu_0\mu^2_{13}\beta_{BB})}H,
\end{eqnarray}
$$G=N^2\sigma_{14}\mu_0\mu^2_{13}\beta_{BE}\beta_{EB}.$$

The susceptibility is a response function of the medium due to an
applied electric field. The electric and magnetic polarizations in
the form of chiral-based electric and magnetic susceptibility are
defined by $P=\epsilon_0\chi_e E+\frac{\xi_{EH}}{c}H$ and
$M=\frac{\xi_{HE}}{\mu_0c}E+\chi_mH$, respectively. These electric
and magnetic susceptibilities for the hot atomic system is written
as:
\begin{eqnarray}
\chi_e(kv)=&\frac{N^2\sigma_{14}\mu_0\mu^2_{13}\beta_{BE}\beta_{EB}+N\sigma^2_{14}\beta_{EE}(\hbar-N
\mu_0\mu^2_{13}\beta_{BB})}{\epsilon_0\hbar(\hbar-N
\mu_0\mu^2_{13}\beta_{BB})}
\end{eqnarray}
and
\begin{eqnarray}
\chi_m(kv)=\frac{N\mu_0\mu^2_{13}\beta_{BB}}{\hbar-N\mu_0\mu^2_{13}\beta_{BB}},
\end{eqnarray}
respectively. The associated chiral coefficients are presented in
the following single equation:
\begin{eqnarray}
\xi_{EH,(HE)}(kv)=\frac{Nc\sigma_{14}\mu_0\mu_{13}\beta_{EB,(BE)}}{(\hbar-N
\mu_0\mu^2_{13}\beta_{BB})}.
\end{eqnarray}
Here, the electric dipole moment is defined as
$\sigma_{14}=\sqrt{3\gamma_1\epsilon_0\hbar c^3/2\omega^3_{14}}$.
We select $v=0$ when the medium is cold and obviously the Doppler
broadening effect can be minimized. The medium is then called as a
cold chiral medium. Nevertheless, If a hot medium is considered,
where the Doppler shift is dominant. we denote it as a hot chiral
atomic medium. For cold atomic medium, we represent
susceptibilities by $\chi_e$ and $\chi_m$, while for the chiral
these are $\xi_{EH}$ and $\xi_{HE}$, respectively. These results
are obtain from Eqs. (18)-(21], when one put $v=0$. The Doppler
susceptibilities are the average of $\chi_e(kv)$, and
$\chi_m(kv)$, over the Maxwellian distribution. The averaged
electric and magnetic susceptibilities can be estimated separately
from the following combined expression
\begin{eqnarray}
\chi_{e,(m)}^{(d)}=\frac{1}{ V_D\sqrt{\pi }}\int^\infty_{-\infty}
\chi_{e,(m)}(kv) e^{-\frac{(kv)^2}{V^2_D}} d(kv)
\end{eqnarray}
where $V_D=\sqrt{K_BT\omega^2Mc^2}$, is the Doppler width. Here,
$\chi_e^{(d)}$ and $\chi_m^{(d)}$ are the Doppler broadened
electric and magnetic susceptibilities for hot atomic system. In
the similar way we can estimate the chiral coefficient terms of
the coupled electric -magnetic fields averaged over the Maxwellian
distribution of the atomic velocity as
\begin{eqnarray}
\xi_{EH,(HE)}^{(d)}=\frac{1}{ V_D\sqrt{\pi }}\int^\infty_{-\infty}
\xi_{EH}(kv)[\xi_{EH}(kv)] e^{-\frac{(kv)^2}{V^2_D}} d(kv)
\end{eqnarray}
We know that the refractive index of a medium is
$n_r=\sqrt{\epsilon\mu}$, where $\epsilon=1+\chi_e$ and
$\mu=1+\chi_m$. The terms $\xi_{EH}$ and $\xi_{HE}$ is the
conurbation from chirality to the refractive index. The
corresponding chiral dependent refractive then gets its shape as:
\begin{eqnarray}
n_{r}&=&Re[((1+\chi_e)(1+\chi_m)-\frac{1}{4}(\xi_{EH}+\xi_{HE})^2)^{1/2}
      \nonumber\\&&+\frac{i}{2}(\xi_{EH}-\xi_{HE})]
\end{eqnarray}
The group refractive index and Group velocity are also presented
here as:
\begin{equation}
N_g=Re[n_r+(\omega_{14}-\Delta_p)\frac{\partial
n_r}{\partial\Delta_p}],
\end{equation}
and
\begin{equation}
v_g=\frac{c}{Re[n_r+(\omega_{14}-\Delta_p)\frac{\partial
n_r}{\partial\Delta_p}]},
\end{equation}
respectively. The corresponding time delay/advance is defined as $
\tau_d=\frac{L(N_g-1)}{c}$. These are the main results which will
be analyzed and discussed in details in the last section.

Further we used the transfer function for observation of out put
pulse shape. The output pulse shape $E_{out}(\omega)$, after
propagating through the medium can be related to the input pulse
shape $E_{in}(\omega)$ by the expression
$E_{out}(\omega)=H(\omega)E_{in}(\omega)$, where
$H(\omega)=e^{-ik(\omega)L}$ is the transfer function for totally
transmitting medium.  we select a Gaussian input pulse of the
form:
\begin{eqnarray}
E_{in}(t)=\exp[-t^2/\tau^2_0]\exp[i(\omega_0+\delta)t],
\end{eqnarray}
where $\delta$, is the upshifted frequency from the empty cavity.
The Fourier transforms of this function is then written by
$E_{in}(\omega)=\frac{1}{\sqrt{2\pi}}\int^{\infty}_{-\infty}E_{in}(t)e^{i\omega
t}d t$. The input signal is calculated as
\begin{eqnarray}
E_{in}\left( \omega \right)  &=&\tau _{0}/\sqrt{2}\exp \left[
-\left( \omega -\omega _{0}-\delta \right) ^{2}\tau^2.
_{0}/4\right]
\end{eqnarray}

The output $E_{out}(t)$ can be obtain from the input pulse by the
convolution theorem as:
\begin{eqnarray}
E_{out}(t)=\frac{1}{\sqrt{2\pi}}\int^{\infty}_{-\infty}E_{int}(\omega)H(\omega)e^{i\omega
t}d\omega
\end{eqnarray}
For simplicity the output pulse shape  is calculated in to the
first order derivative of group index. It is reported as:
\begin{eqnarray}
E_{out}(t)&=&\frac{\tau_0\sqrt{c}}{\sqrt{2iLc
G_{vd}+c\tau^2_0}}\exp[\frac{[2i(L n_0-ct)-\delta\tau^2_0
c]^2}{4c(2iLc
G_{vd}+c\tau^2_0)}]\nonumber\\&&\times\exp[i(t-\frac{n_0L}{c})\omega_0-\frac{\delta^2\tau^2_0}{4}]:
\end{eqnarray}
The inverse fourier transform of $E_{out}(t)$ is $E_{out}(\omega)$
and is written by:
\begin{eqnarray}
E_{out}(\omega)&=&\frac{\sqrt{2}\pi}{\Delta
w}\exp[-\frac{Q_1^2}{4c(2iLcG_{vd}+\frac{4c\pi^2}{(\Delta
w)^2})}]\nonumber\\&&\times\exp[\frac{1}{4}(-\frac{4\pi^2\delta^2}{\Delta
w}+Q_2-\frac{4i\omega_0Ln_0}{c})],
\end{eqnarray}
where
\begin{eqnarray}
Q_1=2iL(\omega-\omega_0)cG_{vd}+\frac{4c\pi^2(\omega-\omega_0-\delta)}{(\Delta
w)^2}+2iL n_0
\end{eqnarray}
and
\begin{eqnarray}
Q_2=\frac{[\frac{4c\pi^2\delta}{(\Delta
w)^2}-2iLn_0]^2}{c(2iLcG_{vd}+\frac{4c\pi^2}{(\Delta w)^2})}.
\end{eqnarray}
In Eqs. (32)-(35), $n_0=\lim_{\omega\longrightarrow\omega_0}N_g$,
$G_{vd}=\lim_{\omega\longrightarrow\omega_0}\frac{\partial
N_g}{c\partial\omega}$ while $\tau_0$ appeared for the input pulse
width in the time domain. Also, $G_{vd}$ is the group velocity
dispersion and $\omega_0$ is the central frequency of the pulse.
Furthermore, the resonances appears at the location
$\omega_0=\omega_{14}$. It is the frequency of the pulse at which
the delay or advancement time is maximum and the distortion is
minimum.

\begin{figure}[t]
\centering
\includegraphics[width=3in]{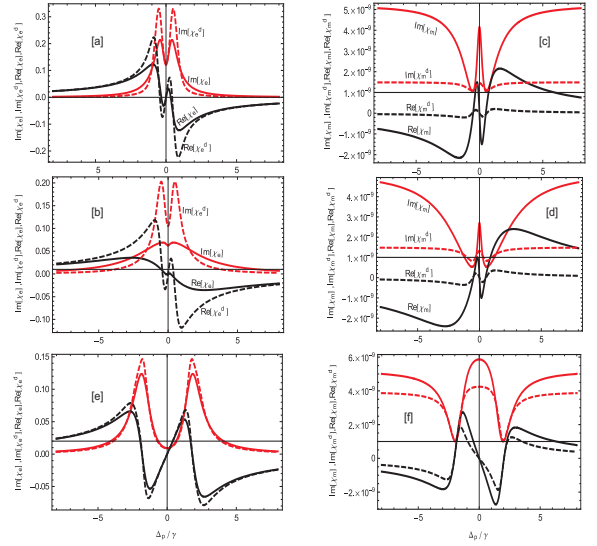}
\caption{ Electric and magnetic Real as well as imaginary
susceptibilities vs probe detuning $\Delta_p/ \gamma$, such that
$\gamma =1GHz$, $\gamma_{1}=\gamma_{1}=
\gamma_{3}=\gamma_{4}=0.1\gamma$, $ \Delta_1=0\gamma$,
$\Delta_2=0\protect\gamma$, $\Delta_b=0\gamma$, $
\Omega_1=0.1\protect\gamma$, $\Omega_2=1\gamma$, $V_D=0.5\gamma$,
$\alpha_i=\pm1$, $\mu_{13}=0.000053\sigma_{14}ms^{-1}$,
$\lambda=1nm$[a,c] $\Omega_3=0.7\gamma$[b,d] $\Omega_3=1\gamma$,
 $\varphi=\pi/2$ [e,f]$\Omega_2=4\gamma$, $\Omega_3=0.7\gamma$, $V_D=0.1\gamma$}
 \label{figure1}
\end{figure}

\begin{figure}[t]
\centering
\includegraphics[width=3in]{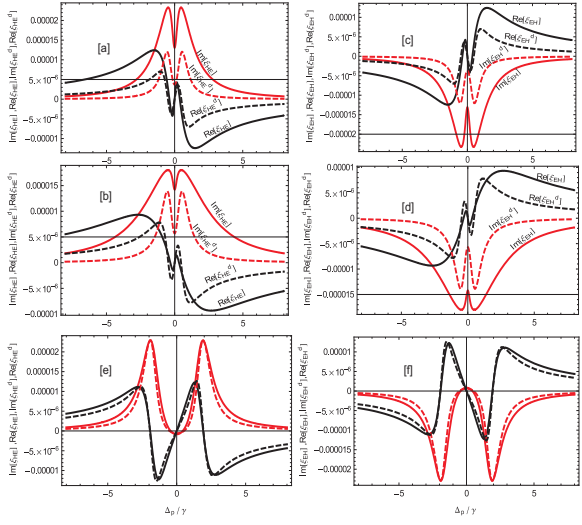}
\caption{   Real and imaginary parts of chiralities vs probe
detuning $\Delta_p/ \gamma$, such that $\gamma =1GHz$,
$\gamma_{1}=\gamma_{1}= \gamma_{3}=\gamma_{4}=0.1\gamma$, $
\Delta_1=0\gamma$, $\Delta_2=0\protect\gamma$, $\Delta_b=0\gamma$,
$ \Omega_1=0.1\protect\gamma$, $\Omega_2=1\gamma$,
$V_D=0.5\gamma$, $\alpha_i=\pm1$,
$\mu_{13}=0.000053\sigma_{14}ms^{-1}$, $\lambda=1nm$[a,c]
$\Omega_3=0.7\gamma$[b,d] $\Omega_3=1\gamma$, $\varphi=\pi/2$
[e,f]$\Omega_2=4\gamma$, $\Omega_3=0.7\gamma$, $V_D=0.1\gamma$}
\label{figure1}
\end{figure}
\begin{figure}[t]
\centering
\includegraphics[width=3in]{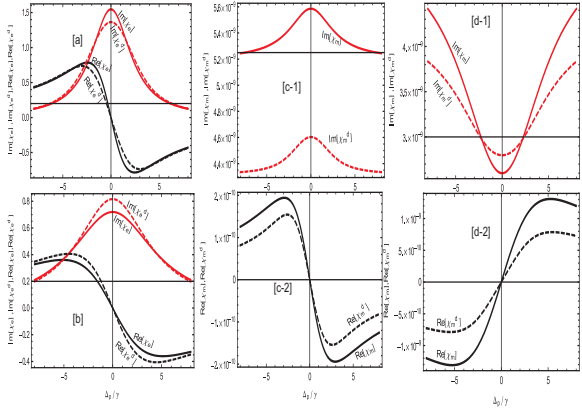}
\caption{ Electric and magnetic Real as well as imaginary
susceptibilities vs probe detuning $\Delta_p/ \gamma$, such that
$\gamma =1GHz$, $\gamma_{1}=\gamma_{1}=
\gamma_{3}=\gamma_{4}=2\gamma$, $ \Delta_1=0\gamma$,
$\Delta_2=0\protect\gamma$, $\Delta_b=0\gamma$, $
\Omega_1=2\protect\gamma$, $\Omega_2=2\gamma$, $V_D=1.5\gamma$,
$\alpha_i=\pm1$,  $\mu_{13}=0.000053\sigma_{14}ms^{-1}$,
$\lambda=1nm$[a,c] $\Omega_3=1.5\gamma$[b,d] $\Omega_3=5\gamma$,
$\varphi=\pi/2$} \label{figure1}
\end{figure}
\begin{figure}[t]
\centering
\includegraphics[width=3in]{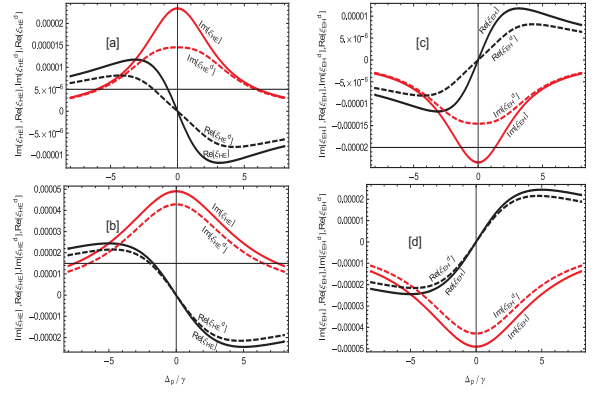}
\caption{ Real and imaginary parts of chirality vs probe detuning
$\Delta_p/ \gamma$, such that $\gamma =1GHz$,
$\gamma_{1}=\gamma_{1}= \gamma_{3}=\gamma_{4}=2\gamma$, $
\Delta_1=0\gamma$, $\Delta_2=0\protect\gamma$, $\Delta_b=0\gamma$,
$ \Omega_1=2\protect\gamma$, $\Omega_2=2\gamma$, $V_D=1.5\gamma$,
$\alpha_i=\pm1$,  $\mu_{13}=0.000053\sigma_{14}ms^{-1}$,
$\lambda=1nm$[a,c] $\Omega_3=1.5\gamma$[b,d] $\Omega_3=5\gamma$,
 $\varphi=\pi/2$} \label{figure1}
\end{figure}

\begin{figure}[t]
\centering
\includegraphics[width=3in]{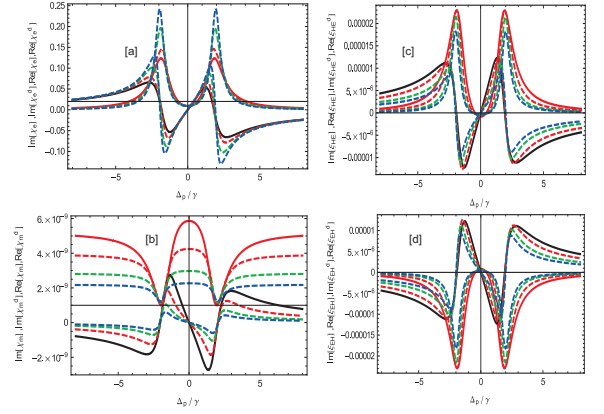}
\caption{ Electric and magnetic Real as well as imaginary
susceptibilities vs probe detuning $\Delta_p/ \gamma$, such that
$\gamma =1GHz$, $\gamma_{1}=\gamma_{1}=
\gamma_{3}=\gamma_{4}=0.1\gamma$, $ \Delta_1=0\gamma$,
$\Delta_2=0\protect\gamma$, $\Delta_b=0\gamma$, $
\Omega_1=0.1\protect\gamma$, $\alpha_i=\pm1$,
$\mu_{13}=0.000053\sigma_{14}ms^{-1}$, $\lambda=1nm$,
 $\varphi=\pi/2$, $\Omega_2=4\gamma$, $\Omega_3=0.7\gamma$, $V_D=0\gamma$(solid), $0.1\gamma$(red dashed),
 $0.2\gamma$(green dashed), $0.3\gamma$(blue dashed)}
 \label{figure1}
\end{figure}

\begin{figure}[t]
\centering
\includegraphics[width=3in]{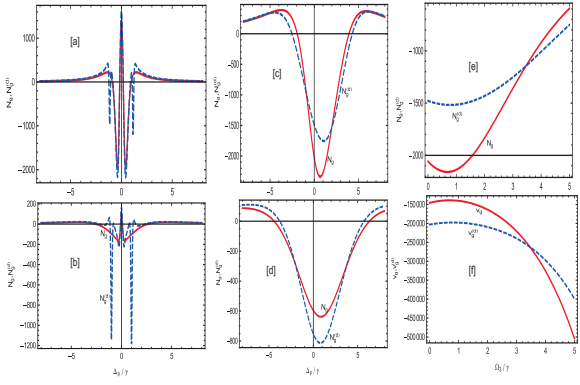}
\caption{ [a,b,c,d] Group index vers probe detuning $\Delta_p/
\gamma$, such that $\gamma =1GHz$, $\gamma_{1}=\gamma_{1}=
\gamma_{3}=\gamma_{4}=2\gamma$, $ \Delta_1=0\gamma$,
$\Delta_2=0\protect\gamma$, $\Delta_b=0\gamma$, $
\Omega_1=2\protect\gamma$, $\Omega_2=2\gamma$, $V_D=1.5\gamma$,
$\alpha_i=\pm1$,  $\mu_{13}=0.000053\sigma_{14}ms^{-1}$,
$\lambda=1nm$, $\omega_{14}=10^4\gamma$,$L=6cm$[a,c]
$\Omega_3=1.5\gamma$[b,d] $\Omega_3=5\gamma$,
 $\varphi=\pi/2$[e,f] Group index and group velocity vers $\Omega_3/\gamma$, using the same parameters but $\Delta_p=0\gamma$} \label{figure1}
\end{figure}
\begin{figure}[t]
\centering
\includegraphics[width=3in]{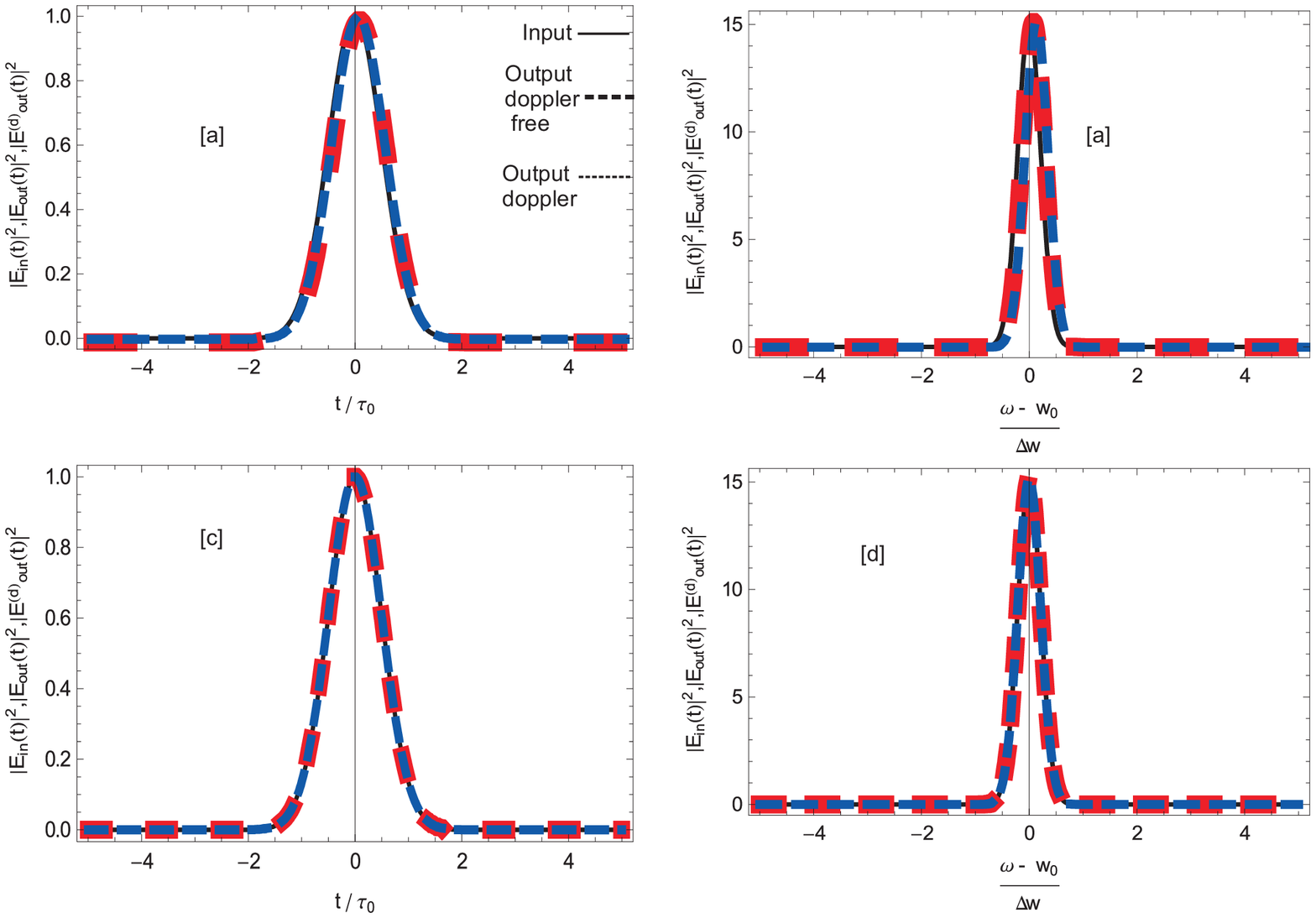}
\caption{[a,b]The normalized Gaussian pulse intensity of input and
outputs at the parameters given Fig2, Vs $\frac{t}{\tau_0}$ and
$\frac{\omega-\omega_0}{\Delta w}$, $\delta=2 GHz $, $\tau_0=5.50n
s$, $\Delta w=2\pi/\tau_0$, $c=3\times10^{8}m/s$, $L=0.06m$,
$n_0=1415.65$, $n^d_0=1618.15$, $G_{vd}=759.44/c$,
$G^{d}_{vd}=18.92/c$, $\omega_0=\omega_{14}=10^4\gamma$[c,d]The
normalized Gaussian pulse intensity of input and outputs at the
parameters given Fig4, Vs $\frac{t}{\tau_0}$ and
$\frac{\omega-\omega_0}{\Delta w}$, $n_0=-2023.81$,
$n^d_0=-1487.22$, $G_{vd}=-9006.67/c$, $G^{d}_{vd}=-344.57/c$}
\label{figure1}
\end{figure}
We explain our main results of the Eqs. (18)-(21)] associated with
the Real and imaginary parts of electric and, magnetic
susceptibilities and their corresponding chiralities, for both
Doppler broadened and Doppler-free systems. The discussion is then
extended to group index, time delay and pulse shape distortion
using their analytical expressions presented in the earlier text.
In Fig. 2, the plots are traced for electric and magnetic
susceptibilities for some spectroscopic parameters. Obviously the
behavior of electric susceptibility is in contrast with  The
electric and magnetic susceptibility. Therefore, the dispersion
associated with electric susceptibility is contrastingly opposite
to the anomalous dispersion of the magnetic one with a less
absorption in comparison. Evidently, the absorption profile
enhances with the hot medium as compared with the cold one.

Intriguingly, the behavior of chirality in case of electric field
coupling is the mirror inversion as compared with the behavior of
magnetic coupling along with an enhanced profile for Doppler-free
system over the Doppler-broadened. Nevertheless, the intensity of
the microwave field generates incoherence processes and disturbs
the interference mechanism of the system. The light propagation in
former corresponds to slow while later it corresponds to fast
light. Amazingly, in the hot medium the Doppler shift develops
coherence and the the dip between the two peak absorption lines
becomes more obvious. Consequently, unlike in the traditional
superluminality where the Doppler shift degrade the super luminal
fast light, here it enhances significantly.

Different spectral behaviors are observed, when there is no
Doppler broadening (cold system) as well as, when there is Doppler
broadening (hot system) in the system. The solid line is for
Doppler free susceptibility and the dashed line is for Doppler
broadened susceptibility. The electric and magnetic
susceptibilities have opposite absorption and dispersion behavior
as shown in  Fig .2[a-d]. The absorption at resonance
$\Delta_p=0$, is reduced in the electric susceptibility, while
there is increase in the magnetic susceptibility. The slope of
dispersion is normal in the electric susceptibility, while
anomalous in the magnetic susceptibility. Nevertheless, the
magnetic effect on absorption and dispersion is very small as
compared with the electric effect on absorption and dispersion.
Further, with the control field $\Omega_3=0.7\gamma$, the
absorption at resonance in both the cases (cold/hot) are $0.2au$.
Nevertheless, the absorption reduces in the Doppler free medium to
$0.05au$ when the control field intensity is increased to
$\Omega_3=1\gamma$ and have small gaps in both the absorption and
dispersion. The modification at $\Delta_p=0$ of slow light
propagation at the resonance point from cold to hot atomic medium
is also obvious. Further, in Fig .3[a-d] the real and imaginary
parts of chiralities $\xi_{HE}$ and $\xi_{EH}$ under similar
condition of Fig 2 are traced which also show opposite behavior
for the real and imaginary parts, respectively.

The real and imaginary parts of chirality $\xi_{HE}$ ($\xi_{EH}$)
shows similarity with electric (magmatic) susceptibility in real
as well as imaginary parts of the susceptibility with inverted
behaviors from cold to hot medium at resonance $\Delta_p=0$ as
shown in Fig 2[c,d] and Fig 3[a,b]. Consequently, their dynamical
characteristics have contrast behavior for the dispersion and
absorption for the two cases, respectively. [see Fig 2[a,b] and
Fig 3[c,d]]. The dominant chiral effect on magnetic susceptibility
is also obvious as compared with the electric one.

Curiositly, controlling the intensity associated with $\Omega_2$,
the absorption in the electric susceptibility and chirality
$\xi_{HE}$ becomes vanished while it is increased for the magnetic
susceptibility and chirality $\xi_{EH}$. The symmetries in
$\chi_e$ and $\xi_{HE}$, and in $\chi_m$ and $\xi_{EH}$ is ideal.
These interesting results are observed at the parameters
$\gamma_{i=1,2,3,4}=0.1\gamma$, $\Omega_1=0.1\gamma$,
$\Omega_2=4\gamma$, $V_D=0.1\gamma$ and $\Omega_3=0.7\gamma$,
$\alpha_{i}=\pm1$ and $\varphi=\pi/2$, as shown in Fig .2[e,f] and
Fig .3[e,f].

In Figs .(4)-(5) the real and imaginary parts of chiralities
$\xi_{EH}$ and $\xi_{EH}$, are shown with the same parameters. The
absorption of cold medium is less (large) than the absorption of
hot medium at low (large) intensity of the control field
$\Omega_3=1.5\gamma$, at the resonance point. The real and
imaginary parts of the chirality $\xi_{HE}$ have similar spectral
behavior as compared with $\chi_e$ for low intensity of the
control field for both the cold and hot atomic medium.
Nevertheless, for high intensity of the control field the spectral
behaviors are inverted for both the cold and hot atomic medium as
shown in Fig 4[a,b] and Fig 5[a,b]. Furthermore, similar
characteristics is also true for $\xi_{EH}$ for its real and
imaginary parts of the magnetic when the the intensity of the
control field is high as shown in Fig 4[d] and Fig 5[d]. In Fig .6
the electric and magnetic susceptibilities and the chiralities are
traced with the Doppler widths, a parameter controllable with
temperature of the medium. A very dominant response is seen about
the time delay and advance when the Doppler width is increased
stepwise. Consequently, going cold chiral medium to a hot chiral
medium the behavior of the system can be reverted.

The group index and time delay/advancement for the parameters of
Fig .2 is shown in Fig. 7.  At the resonance point
$\Delta_p/\gamma=0$ sharp positive peaks of group indices and
times delays/advancement are observed. When the intensity of the
control field $\Omega_3$ is $0.5\gamma$, the value of group index
for cold medium is $N_g=1415.65$ and the value of group index for
hot medium is $N^{(d)}_g=1618.15$. The corresponding group delays
are $t_d=121n s$ and $t^{(d)}_d=917n s$. If the intensity of
control field $\Omega_3$ is increased from $0.7\gamma$ to
$1\gamma$. Then the group index are $N_g=110.96$ and
$N^{(d)}_g=164.013$, while delays in times are $t_d=219 ns$ and
$t^{(d)}_d=997 ns$, respectively. Correspondingly, the group
velocities vary from $v_g=c/1415.65$ to $v_g=c/110.96$ and
$v^{(d)}_g=c/1618.15$ to $ v^{(d)}_g=c/164.013$. The above results
are spatially important to the preservation of two light pulses.
When two light pulses reach at same times to the instrument which
store its information. The instrument stored the information of
one of the pulse and lost the information of the other pulse.
There delays in times recovered this difficulty and the
information of both the pulses will be preserved. In Fig. 7 (c,d),
for $\Delta_p=0\gamma$, the values of group indices are
$N_g=-2023.81$ and $N^{(d)}_g=-1487.22$ at the intensity of the
control field $\Omega_3=1.5\gamma$. The corresponding advance
times are $t_{ad}=-404 ns$ and $t^{(d)}_{ad}=-297 ns$. In this
case the cold medium is more superluminal, and the advance time is
larger then the hot medium by $107 ns$. When the intensity of the
control field is increase from $1.5\gamma$ to $5\gamma$, the group
index of the cold medium is $N_g=-595.818$, while the group index
of the hot medium is $N^{(d)}_g=-751.666$. The corresponding
advance times are then $t_{ad}=-119.36 s$ and
$t^{(d)}_{ad}=-150.53n s$. The advance time difference is $31.17
ns$. In this case the Doppler broadened medium is more
superluminal than the Doppler free medium.

In the previous related studies, it has been observed that the
Doppler broadened medium is less superluminal, than a Doppler free
medium. Here a chiral medium shows contrastingly different
behavior, that it may enhance superluminality at certain
conditions of the coherent control field shown in Fig 7[d]. This
fact is more clearly observed in the Fig 7[e,f]. It is clear from
these plots that below the intensity of the control field
$\Omega_3=3.6\gamma$, the group index of cold chiral medium is
more negative with a larger group velocity as compared with the
hot chiral medium. Therefore the chiral cold medium is more
superluminal than the chiral hot medium below the intensity of
control field of $3.6\gamma$. Nevertheless, for
$\Omega_3>3.6\gamma$, the group index of the chiral hot medium is
more negative and larger negative group velocity as compared to
the chiral cold medium, Hence chiral hot medium is more
superluminal above the intensity of the control field of
$3.6\gamma$. Furthermore, at high intensity of the control field
the group index and group velocity becomes saturated (not shown).

Fig .8 shows Gaussian pulse shapes of input and output of the
chiral medium. All the pulses are the same shapes for cold and hot
medium, if the coherent fields counter propagate to the probe
field or co-propagate to the probe field up to first order group
velocity dispersion. The pulses are fully distortion-less to the
first order group index derivative with respect to angular
frequency. However if one counts higher order derivative then the
pulse is slowly distorted.

In conclusion, we study Chiral Based Electromagnetically Induced
Transparency (CBEIT) of a light pulse and its associated
subluminal and superluminal behavior through a cold and a hot
medium of 4-level \textit{double-Lambda type} atomic system. The
dynamical behavior of this chiral based system is temperature
dependent. The magnetic field based chirality and dispersion is
always opposite as compared with the electric field ones.
Contrastingly, the response of the chiral effect along with the
incoherence Doppler broadening mechanism enhances the superluminal
behavior as compared with its traditional degrading effect.
Nevertheless, the intensity of a coupled microwave field destroys
the coherence of the medium and degrade superluminality and
subluminality of the sysmtem. The undistorted retrieved pulse from
a hot chiral medium delays by $896 ns$ than from a cold chiral
medium under same set of parameters. Nevertheless, it advances by
$-31n s$ in the cold chiral medium when a suitably different
spectroscopic parameters are selected. The corresponding group
index of the medium and the time delay/advance, are studied and
analyzed explicitly [Note: A revise version is under preparation

\end{document}